# A Nuclear Magnetic Resonance Study on Rubrene-cobalt Nano-composites

Masashi Shiraishi[1,2], Haruka Kusai[1], Ryo Nouchi[1], Takayuki Nozaki[1], Teruya Shinjo[1] and Yoshishige Suzuki[1]

1. Graduate School of Engineering Science, Osaka University,
   Machikaneyama-cho 1-3, Toyonaka 560-8531, Osaka, Japan
2. PRESTO-JST, Honmachi 4-1-8, Kawaguchi 332-0012, Saitama, Japan

Makoto Yoshida[3] and Masashi Takigawa[3]

3. Institute for Solid State Physics, The University of Tokyo, Kashiwa, Chiba 277-8581, Japan

**Abstract**

We implemented a nuclear magnetic resonance (NMR) study on rubrene($C_{42}H_{28}$)-Co nano-composites that exhibit an enhanced magnetoresistance (MR) ratio of 80%. The $^{59}$Co NMR spin echo experiment enabled clarification of the hyperfine field of Co at the interface between the ferromagnet and the molecules, which has not been investigated for molecular spintronics. An enhanced hyperfine field of the Co was observed in the rubrene-Co nano-composites, which may be related to the enhancement of the MR ratio. This study demonstrates the importance of microscopic investigation of the interface between molecules and ferromagnets that governs spin-dependent transport in molecular spin devices.

Molecule (MO)-ferromagnet (FM) nano-composites [1-6], where FM nano-particles are embedded into a MO matrix similar to a granular system (Co particles in a Cu matrix [7,8]), have attracted much attention in recent years because they exhibit very large magnetoresistance (MR) effects at low temperatures. [2-5] The maximum MR ratio typically ranges from 80% in rubrene($C_{42}H_{28}$)-Co [4] to 400% in $C_{60}$-Co [6] at 4.2 K, dramatically exceeding the theoretically-predicted values based on a Julliere model. These values are also larger than that in inorganic granular systems, such as Co-Al-O. [9] Furthermore, it has been verified that the MR effect was induced by magnetization of the Co particles. [2-5] Such a large MR ratio in MO-NM nano-composites is so attractive for practical applications of molecular spin devices. For example, spin transistors and magnetic sensors may be fabricated if the MR ratio is large enough at room temperature (RT). Currently, the large MR ratio is limited to low temperatures, while the maximum MR at RT is only 0.3 % among all nano-composite devices. We have described the importance of a $C_{60}$/Co interface in the temperature dependence of the MR ratio. [3] These studies have been motivated by the establishment and development of a field of molecular spintronics in which various studies have been implemented. Although a hysteresis in resistance was observed, there have been very few detailed and precise experiments, which would facilitate the understanding of underlying physics and clarify the origin of spin-dependent transport. Furthermore, a MO/FM interface has been characterized in molecular spin devices by a limited number of studies. [10-12] Thus, most of the studies on molecular spintronics exploiting 2-terminal structures have been controversial. [13,14]

Thus far, the origin of the enhanced MR ratio in MO-FM nano-composites have been explained in two ways; the charging effect, [9] which is induced by Coulomb blockade in spin transport via FM nano-particles, and enhancement of spin polarization at the MO/FM interface. [15] However, this issue remains unclear. Nuclear magnetic resonance (NMR) is a powerful tool for clarification of the surface and interface spin states in FM materials, because the hyperfine field is highly sensitive to the local surroundings. [16] For instance, $^{59}$Co NMR experiments have enabled the clarification of spin states of a Co thin film [17], Co/Cu artificial lattice [18] and Co-Cu granular [19,20]. Hence, an NMR study is an appropriate tool to carry out characterization of the interface between MO and FM in order to examine the possibility for the enhancement of the spin polarization of the Co in the nano-composites, which thus far has not been investigated. Whilst the contribution of the charging effect to the large MR ratio has been described elsewhere, [21] we herein report on $^{59}$Co NMR in rubrene-Co nano-composites which exhibited a maximum MR ratio of 80% at 4.2 K. [3] Furthermore, we discuss the origin of the enhanced MR ratio from the point of the interface spin state between rubrene and Co.

We have prepared three different samples on polyethylene naphthalate (PEN) substrates; a Co thin film (350 nm in thick), a rubrene-Co nano-composite (content ratio=3:1, 1400 nm in thick) and a rubrene film (1050 nm in thick). A glass-capping layer was evaporated onto every sample in order to prevent oxidation of the Co nano-particles and the rubrene. A co-evaporation method was adopted in order to synthesize the rubrene-Co nano-composites, in which the mean diameter of the Co nano-particles was estimated to be 0.8-2.3 nm based on our previous report. [4] Hence, the magnetic structure of the Co nano-particles is most likely a single domain and more than a half of the Co atoms occupy surface positions. After the fabrication, the samples were cut to 50 mm$^2$ (in total 48 pieces) and placed in an NMR spectrometer. The measuring temperature was 2 K. The NMR signal was observed by a spin echo spectrometer with coherent detection using a tuned circuit matched to a 50 Ω line. The NMR spectra were obtained by recording the peak height of the spin echo signal at discrete frequencies in the range between 100 and 320 MHz. We have not made any correction for the signal intensity to compensate for variation of the nuclear polarization or the sensitivity of the spectrometer over the measured frequency range.

The upper panel of Fig. 1 displays a spin echo spectrum from the Co thin film. Two characteristic peaks from the different crystal structures of Co can be seen at 217 and 224 MHz, which correspond to fcc- and hcp-Co, respectively. These peaks shifted to lower frequency by $\gamma H$ as a magnetic field $H$ was applied (Zeeman shift), where $\gamma$ (=10.1021 MHz/T) is the nuclear gyromagnetic ratio. The peak at 224 MHz was also observed in the nano-composite samples, which is likely due to the hcp-Co. However, it is noted that a spin echo spectrum in the rubrene-Co nano-composite was very different from that of the Co film (lower panel in Fig. 1). Here, it was verified that no signal was observed in the rubrene thin film and the signals from the rubrene-Co nano-composite are due to the Co nano-particles in the sample. First, several peaks were observed in the frequency range below 220 MHz, which is likely ascribed to microscopic changes in the environmental conditions around the Co atoms in the nano-particles. The appearance of several peaks with lower frequencies is understood in analogy with results obtained for the Co/Cu artificial lattice systems [18]; namely a small number of impurities located in neighboring positions of the Co atoms due to co-evaporation with rubrene may give rise to a downshift in frequency. Second, the spin echo signals at a higher frequency ranging from 230 MHz to 260 MHz were observed, which indicates an enhanced hyperfine field in comparison with that in the pure Co film. Similar enhancement of the hyperfine field of Co was also reported in the Co-SiO$_2$ granular, [22] but no enhancement was observed in the Co-Cu granular alloy. [23] Although the authors of ref. 22 attributed this enhanced hyperfine field to a clustering effect, we infer that it may originate from Co sites at the interface between the Co nano-particles and the rubrene. [24] The hyperfine field at Co nuclei is due to polarization of inner core s-electrons, which is proportional to the polarization of d-electrons on the same site. Hence the enhanced hyperfine field implies larger magnetic moment. Even if the enhanced hyperfine field is due to larger magnetic moment at the interface, the enhancement is only 10 %, which is not sufficient to explain the observed large MR ratio. In any case, our finding provides important information to understand the spin states at the interface.

Measured spectra of the nano-composite under application of an external magnetic field up to 6 T are displayed in Fig. 2. All spectra are upshifted by $\gamma H$ to compensate for the Zeeman shift due to the external field. As can be seen, the dominant peak at 224 MHz is diminished and fine patterns of the spectrum at 0 T were not reproduced under application of the external magnetic field, whereas the spectrum in the frequency rage higher than 235 MHz was not affected by the field. This unique behavior could be interpreted in two ways: (1) The peak at 244 MHz presumably comes from those Co sites of which local coordination is similar to that in hcp Co. Therefore, the hyperfine field is expected to be anisotropic. As the magnetization of the Co nano-particles is aligned parallel to the magnetic field, the relative orientation of the magnetization with respect to the crystal axis of each particle becomes random. The spectrum should then be broadened by the anisotropy of the hyperfine field. (2) The magnetization may not be completely saturated even in the fields up to 6 T. Since the resonance frequency is given by the vector sum of the hyperfine field and the external field, the distribution of the orientation of the magnetization relative to the external field should cause broadening of the spectrum. However, since the magnetization of a similar sample of Co nano-particles measured by SQUID reached more the 80% of saturation at the field of 1 T [4], the second possibility is less likely.

We have measured the spin relaxation time ($T_2$). Under a zero magnetic field, two component processes were observed as previously reported. [22] The results are shown in Fig. 3 and Table 1. The fitting was carried out using two exponential functions of the form,

$$A(2\tau) = A(0)\exp(-\frac{2\tau}{T_2}). \qquad (1)$$

The fast relaxation process is thought to be associated with small Co clusters, and the slower process is a conventional relaxation process (spin-spin relaxation). Because the $T_2$ values were not significantly different from each other, it can be concluded that the signal intensity is proportional to the spin number. Under application of the magnetic field, $T_2$ became longer as expected.

In summary, NMR studies were implemented in rubrene-Co nano-composites where an MR ratio of 80% was observed. The nano-composites exhibited higher frequency signals that indicate an enhanced hyperfine field in the Co nano-particles compared with those from a pure Co thin film. We deduce that this hyperfine field would be partially attributed to an enhanced MR ratio in the rubrene-Co nano-composites. Our finding provides important information toward our understanding of the interface spin states at the rubrene/Co interface and the enhancement of the MR ratio in rubrene-Co nano-composites.

This study was supported in part by the Asahi Glass Foundation. The PEN substrates were provided by Teijin-Dupont Film Inc.

# Table captions

**Table 1**

The two-component $T_2$ values at a zero magnetic field. $\tau_I$ and $\tau_{II}$ are shorter and longer spin relaxation time constants, respectively.

|  | 171 MHz | 199 MHz | 224 MHz | 242 MHz |
|---|---|---|---|---|
| $\tau_I$ (μsec) | 14 | 14 | 11 | 22 |
| $\tau_{II}$ (μsec) | 30 | 60 | 52 | 51 |

# Figure captions

**FIG. 1 (color online)**

$^{59}$Co NMR spectra for the Co thin film (upper panel) and for the rubrene-Co nano-composites (lower panel) at 2 K.

**FIG. 2 (color online)**

The magnetic field dependence of the $^{59}$Co NMR spectrum for the rubrene-Co nano-composite. The external magnetic field was set to be 0, 1, 2, 3 and 6 T. All spectra are shifted in proportional to the gyromagnetic ratio, $\gamma$ (=10.1021310 MHz/T)

**FIG. 3 (color online)**

Decay of the $^{59}$Co spin echo with pulse separation at 171, 199, 224 and 242 MHz at 2 K. A fast and a slow relaxation time can be clearly seen.

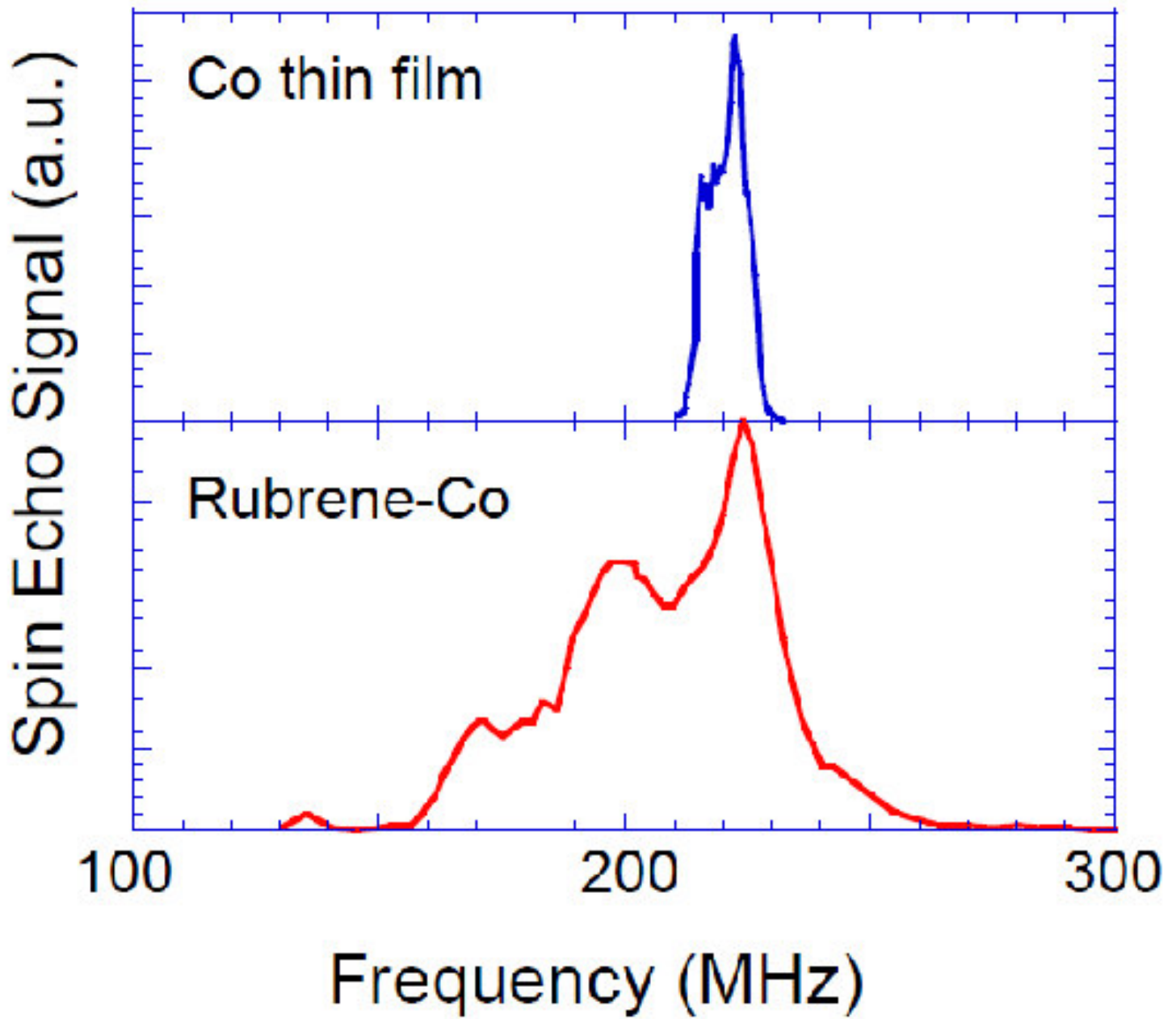

Fig.1 Shiraishi et al.

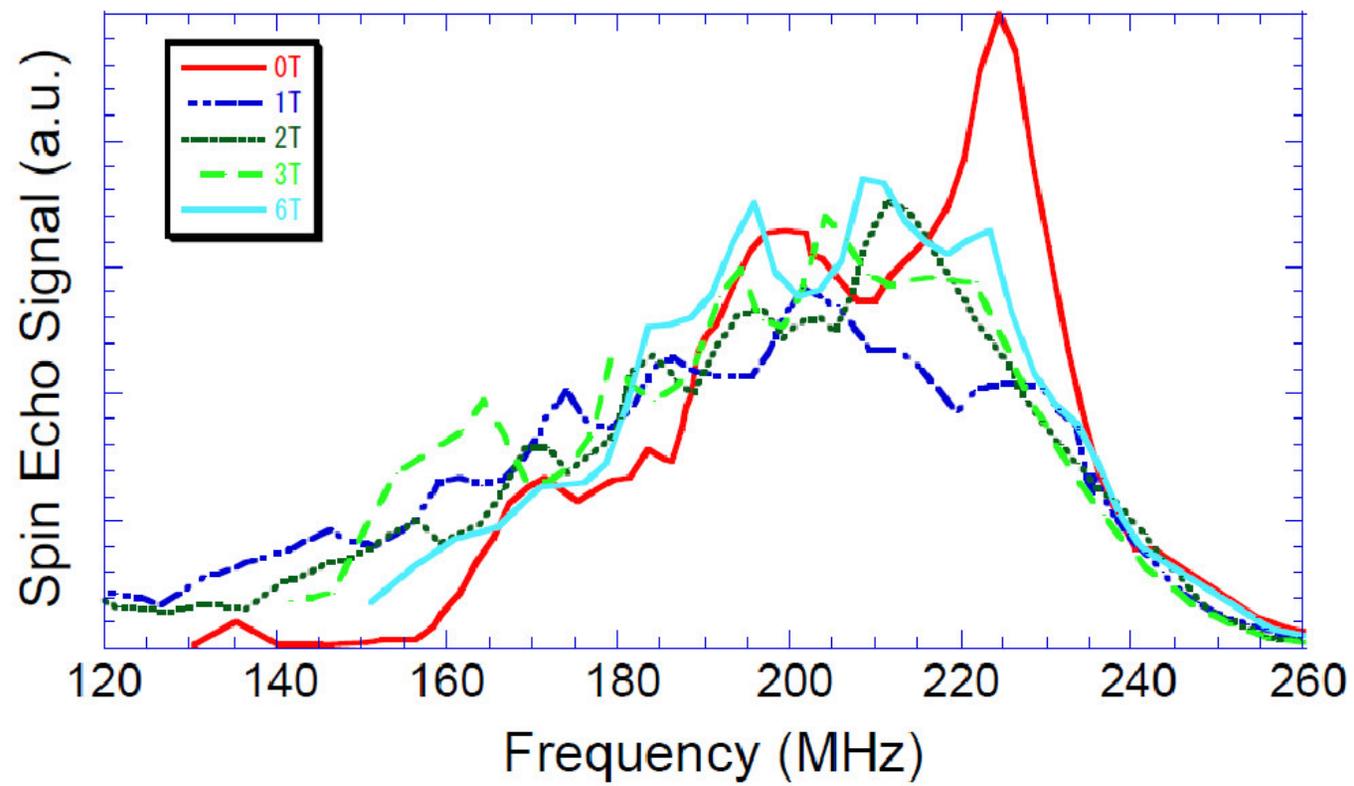

Fig. 2 Shiraishi et al.

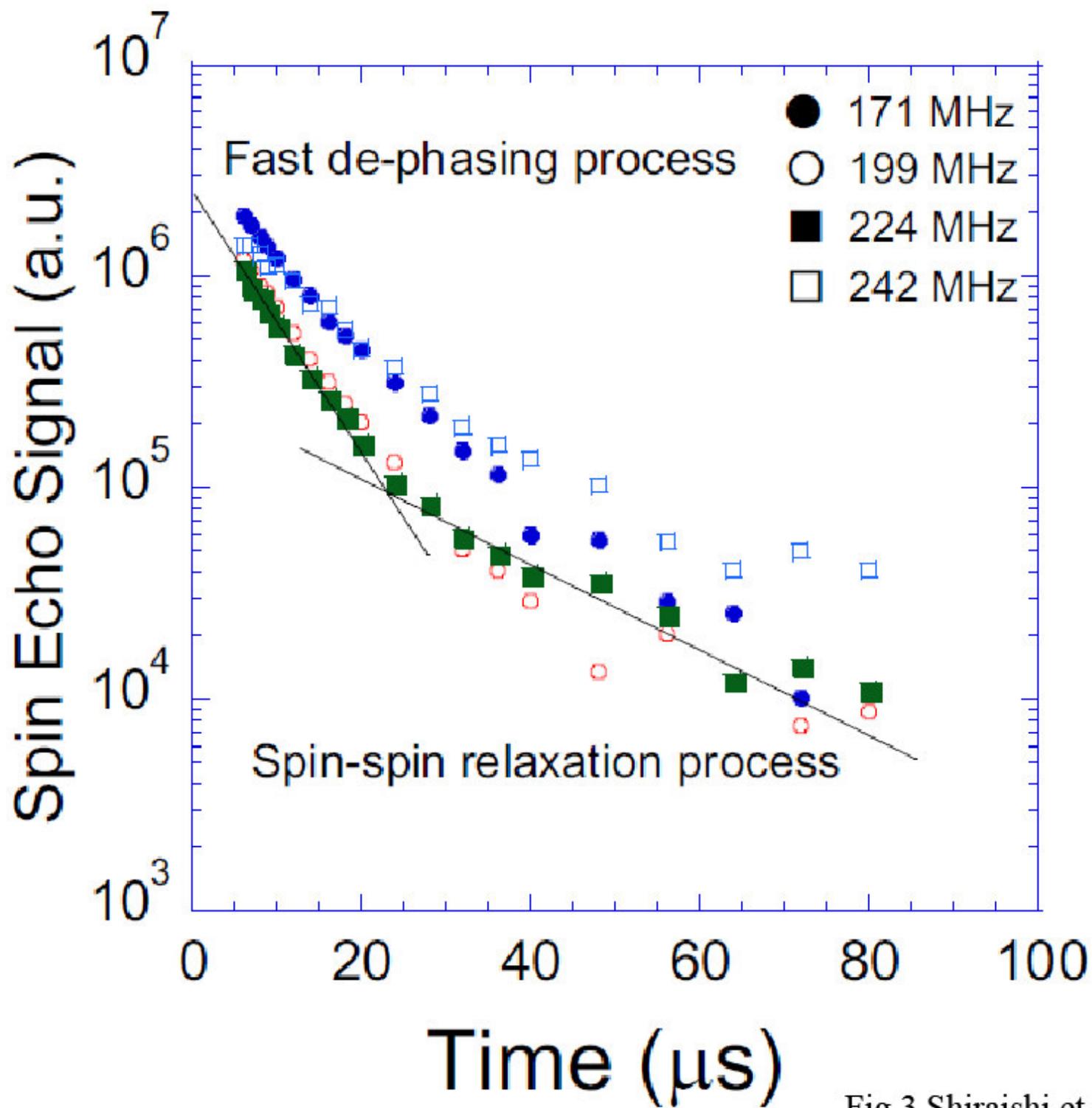

Fig.3 Shiraishi et al.